\documentclass[twocolumn,english,aps,prb,floats]{revtex4}
\usepackage[T1]{fontenc}
\usepackage[latin9]{inputenc}
\usepackage{babel}
\usepackage{amsmath}
\usepackage{amssymb}
\usepackage{graphicx}
\usepackage{esint}
\usepackage[unicode=true,
 bookmarks=false,
 breaklinks=false,pdfborder={0 0 1},backref=section,colorlinks=false]
 {hyperref}

\makeatletter
\@ifundefined{textcolor}{}
{%
 \definecolor{BLACK}{gray}{0}
 \definecolor{WHITE}{gray}{1}
 \definecolor{RED}{rgb}{1,0,0}
 \definecolor{GREEN}{rgb}{0,1,0}
 \definecolor{BLUE}{rgb}{0,0,1}
 \definecolor{CYAN}{cmyk}{1,0,0,0}
 \definecolor{MAGENTA}{cmyk}{0,1,0,0}
 \definecolor{YELLOW}{cmyk}{0,0,1,0}
}

\usepackage{babel}
\usepackage{bm}
\usepackage{wasysym}
\usepackage{breakurl}

\@ifundefined{textcolor}{}{%
 \definecolor{BLACK}{gray}{0}
 \definecolor{WHITE}{gray}{1}
 \definecolor{RED}{rgb}{1,0,0}
 \definecolor{GREEN}{rgb}{0,1,0}
 \definecolor{BLUE}{rgb}{0,0,1}
 \definecolor{CYAN}{cmyk}{1,0,0,0}
 \definecolor{MAGENTA}{cmyk}{0,1,0,0}
 \definecolor{YELLOW}{cmyk}{0,0,1,0}
}

\@ifundefined{textcolor}{}{%
 \definecolor{BLACK}{gray}{0}
 \definecolor{WHITE}{gray}{1}
 \definecolor{RED}{rgb}{1,0,0}
 \definecolor{GREEN}{rgb}{0,1,0}
 \definecolor{BLUE}{rgb}{0,0,1}
 \definecolor{CYAN}{cmyk}{1,0,0,0}
 \definecolor{MAGENTA}{cmyk}{0,1,0,0}
 \definecolor{YELLOW}{cmyk}{0,0,1,0}
}

\usepackage{babel}
\usepackage{ifpdf}\usepackage{bm}

\makeatother

\begin{document}
\title{Absorption of Microwaves by Random-Anisotropy Magnets}
\author{Dmitry A. Garanin and Eugene M. Chudnovsky}
\affiliation{Physics Department, Herbert H. Lehman College and Graduate School,
The City University of New York, 250 Bedford Park Boulevard West,
Bronx, New York 10468-1589, USA }
\date{\today}

\begin{abstract}
Microscopic model of the interaction of spins with a microwave field
in a random-anisotropy magnet has been developed. Numerical results
show that microwave absorption occurs in a broad range of frequencies
due to the distribution of ferromagnetically correlated regions on sizes and effective
anisotropy. That distribution is also responsible for the weak dependence
of the absorption on the damping.  At a fixed frequency of the ac-field
spin oscillations are localized inside isolated correlated regions. Scaling of the peak absorption frequency agrees with the theory based upon Imry-Ma argument. The effect of the dimensionality of the system related to microwave absorption by thin amorphous magnetic wires and foils has been studied. 
\end{abstract}

\maketitle

\section{Introduction}

In conventional ferromagnets the ac field can excite spin waves with
a finite angular momentum and/or induce the uniform ferromagnetic resonance
(FMR) corresponding to the zero angular momentum. In the presence
of strong disorder in the local orientation of spins, however, that exists in materials with quenched randomness, such a spin glasses and amorphous ferromagnets, spin waves must
be localized while the existence of the FMR becomes non-obvious.
On general grounds one should expect that random magnets would exhibit
absorption of the ac power in a broad frequency range that would narrow
down when spins become aligned on increasing the external magnetic field.

Collective excitation modes have been observed in random magnets in
the past \cite{Monod,Alloul1980,Schultz}. In spin-glasses they were
attributed \cite{Fert} to the random anisotropy arising from Dzyaloshinskii-Moriya
interaction and analyzed \cite{Henley1982} within hydrodynamic theory
\cite{HS-1977,Saslow1982}. Later Suran et al. studied collective
modes in amorphous ferromagnets with random local magnetic anisotropy
\cite{Suran-RA} and reported evidence of their localization \cite{Suran-localization}.
Longitudinal, transverse and mixed modes have been observed in thin
amorphous films. Detailed analysis of these experiments, accompanied
by analytical theory of the uniform spin resonance in the random anisotropy
(RA) ferromagnet in a nearly saturating magnetic field, has been recently
given by Saslow and Sun \cite{Saslow2018}.

Rigorous approach to this problem requires investigation of the oscillation
dynamics of a system of a large number of strongly interacting spins
in a random potential landscape. While it was not possible at the
time when most of the above-mentioned work was performed, the capabilities
of modern computers allow one to address this problem numerically
in great detail. Such a study must be worth pursuing because of the absence
of the rigorous analytical theory of random magnets and also with
an eye on their applications as microwave absorbers.

In this paper we consider dynamics of an amorphous ferromagnet consisting
of up to $4 \times 10^5$ spins within the RA model. It assumes (see, e.g.,
Refs. \onlinecite{CSS-1986,CT-book,PCG-2015} and references therein)
that spins interact via ferromagnetic exchange but that directions
of local magnetic anisotropy axes are randomly distributed from one spin to another. In the past this model was successfully applied
to the description of static properties of amorphous magnets, such
as the ferromagnetic correlation length, zero-field susceptibility,
the approach to saturation, etc. \cite{RA-book}.

The essence of the RA model can be explained in the following terms. The ferromagnetic
exchange tends to align the spins in one direction but it has no preferred
direction. In a crystalline body the latter, in the absence of the magnetic field, is determined
by the magnetic anisotropy due to the violation of the rotational
symmetry by the crystal lattice. Still, due to the time reversal symmetry,
any two states with opposite directions of the magnetization have the
same energy. In a macroscopic magnet this leads to the formation of
ferromagnetically aligned magnetic domains. Magnetic particles of size below one micron typically
consist of one such a domain.

\begin{figure}[h]
\centering{}\includegraphics[width=8.5cm]{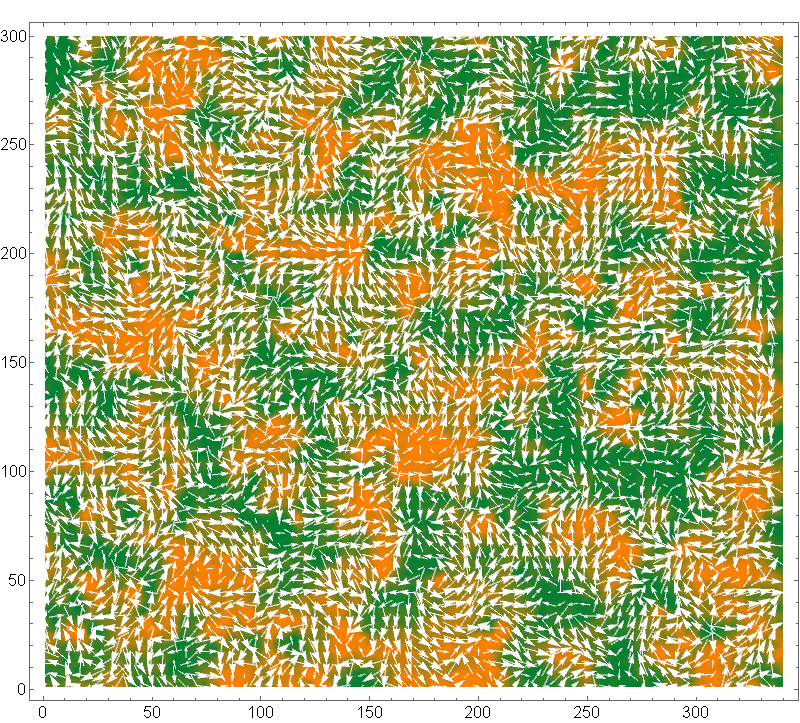} \caption{Spin configuration in a 2D RA ferromagnet obtained by relaxation
from random initial orientations of three-component Heisenberg spins. Spins form correlated regions - 
Imry-Ma domains. The color coding reflects the sign of the out-of-plane $s_z$ component of the spin, with orange/green
corresponding to positive/negative. The in-plane spin components $s_{x}$,
$s_{y}$ are shown by white arrows.}
\label{Fig-IM}
\end{figure}
This changes in an amorphous magnet. If no material anisotropy was introduced
by a manufacturing process, such a magnet would be lacking global anisotropy
axes. Random on-site magnetic anisotropy disturbs the local ferromagnetic
order but cannot break it at the atomic scale $a$ because the RA
energy per site, $D_{R}$, is small compared to the exchange energy
$J$. The resulting magnetic state can be understood within the framework
developed in the seminal papers of Larkin \cite{Larkin} and Imry
and Ma \cite{IM}. Due to random local pushes from the RA the magnetization
wonders around the magnet at the nanoscale in a random-walk manner (see Fig.\ \ref{Fig-IM}), with the ferromagnetic
correlation length given by $R_{f}/a\propto(J/D_{R})^{2/(4-d)}$,
where $d$ is the dimensionality of the system.

This statement, known as the Imry-Ma (IM) argument, works for many
systems with quenched randomness, such as disordered antiferromagnets \cite{Aharony}, flux lattices in superconductors
\cite{Blatter}, charge-density waves \cite{Gruner}, liquid crystals
and polymers \cite{Bellini,Radzihovsky}, and superfluid $^{3}$He-A
in aerogel \cite{Volovik}. In the case of the RA ferromagnet it suggests
that the RA, no matter how weak, breaks the long range ferromagnetic
order for $d=1,2,3$, although the order can persist locally on the
scale $R_{f}$ that can be large compared to the interatomic distance
$a$ if $D_{R}\ll J$. For the same ratio $D_{R}/J$, the lower is
the dimensionality of the system the smaller is the correlated region.
(Formally, the long-range order is restored in higher dimensions,
$d=4,5,...$.)

Fundamental feature of the RA model is that it can be rescaled in
terms of spin blocks of size $r>a$ with the effective $D'_{R}$ and
$J'$ different from the original $D_{R}$ and $J$ up to the size
$r$ at which $D'_{R}\sim J'$. This can be useful for numerical work
but it also means that the model is intrinsically non-perturbative.
It describes a strongly correlated system that cannot be treated perturbatively
on small $D_{R}/J$. The latter is evidenced by the IM result for the ferromagnetic
correlation length $R_{f}$.

One deficiency of the IM model is that it ignores topological defects
\cite{PGC-PRL} (apparent in Fig.\ \ref{Fig-IM}) that lead to metastability.
It was recently argued that random field (RF) converts a conventional ferromagnet
into a topological glass in which ferromagnetically correlated regions
(often called IM domains) possess nonzero topological charges \cite{CG-PRL}.
Although this argument was made for the RF rather than
the RA, the two models have much in common due to the fact that the
RA creates a local anisotropy field that acts on spins similarly to
the RF. The RA model, however, is more nonlinear than the RF model.
High metastability, history dependence, and memory effects \cite{GC-EPJ}
exhibited by ferromagnets with random magnetic anisotropy reveal complex
non-ergodic temporal behavior typical of spin glasses \cite{nonergodic}.

Fueled by potential applications, there has been a large body of recent
experimental research on the absorption of microwave radiation by
nanocomposites comprised of magnetic nanoparticles of various shapes
and dimensions, embedded in dielectric matrices \cite{nanocomposites}.
With the use of more and more exotic shapes and materials, the complexity
of such systems has increased dramatically in recent years \cite{Sun-2011,nanomaterials,carbon}
but their evaluation as microwave absorbers has been largely empirical and often a matter
of luck rather than driven by theory.

Here we investigate the microwave absorption by the RA magnet in a
zero external field - situation the least obvious from the theoretical
point of view and the least studied in experiments, although it must be the most interesting one
for applications. Our goal is to understand the fundamental physics
of the absorption of the ac power by the random magnet without focusing
on material science. The reference to microwaves throughout the paper is determined by
the outcome: The peak absorption happens to be in the microwave range
due to the typical strength of the magnetic anisotropy.

We should emphasize that this problem is noticeably different from
the microwave absorption by a nanocomposite. Coated magnetic particles
or particles dissolved in a dielectric medium are absorbing the ac
power more or less independently up to a weak dipole-dipole interaction
between them. On the contrary, in the amorphous ferromagnet all spins
are coupled by the strong exchange interaction and they respond to
the ac field collectively. Metastability and magnetic hysteresis exhibited
by the RA magnet \cite{PCG-2015} makes such response highly nontrivial.

Our most interesting observation is that the absorption of the microwave
power in the RA magnet is dominated by spin oscillations localized
inside well-separated ferromagnetically correlated regions (IM domains)
that at a given frequency are in resonance with the microwave field.
In that sense there is a similarity with a nanocomposite where certain
particles react resonantly to the ac field of a given frequency. However,
the pure number of such areas in a random magnet must be greater due
to the higher concentration of spins.

The paper is organized as follows. The RA model and the numerical
method for computing the absorption of the ac power are introduced
in Section \ref{model}. Results of the computations are given in
Section \ref{results}. Interpretation of the results, supported by
snapshots of oscillating spins, is suggested in Section \ref{interpretation}.
Estimates of the absorbed power and implications of our findings for
experiments are discussed in Section \ref{discussion}.

\section{The model and numerical method}

\label{model}

We consider Heisenberg RA model described by the Hamiltonian
\begin{equation}
\mathcal{H}=-\frac{J}{2}\sum_{i,j}\mathbf{s}_{i}\cdot{\bf s}_{j}-\frac{D_{R}}{2}\sum_{i}({\bf n}_{i}\cdot{\bf s}_{i})^{2}-\mathbf{h}(t)\cdot\sum_{i}{\bf s}_{i},\label{Hamiltonian}
\end{equation}
where the first sum is over nearest neighbors, ${\bf s}_{i}$ is a
three component spin of a constant length $s$, $D_{R}$ is the strength
of the easy axis RA, ${\bf n}_{i}$ is a three-component unit
vector having random direction at each lattice site, and ${\bf h}$ is the magnetic field in energy units. We assume ferromagnetic
exchange, $J>0$. Factor $1/2$ in front of the first term is needed
to count the exchange interaction $Js^{2}$ between each pair of spins
once. In our numerical work we consider a chain of equally spaced
spins in 1D, a square lattice in 2D, and a cubic lattice in 3D. For
the real atomic lattice of square or cubic symmetry the single-ion
anisotropy of the form $-({\bf n}\cdot{\bf s})^{2}$ would be absent,
the first non-vanishing anisotropy terms would be fourth power on
spin components. However, in our case the choice of the lattice is
merely a computational tool that should not affect our conclusions.

The last term in Eq.\ (\ref{Hamiltonian}) describes Zeeman interaction
of the spins with the ac magnetic field of amplitude ${\bf h}_{0}$
and frequency $\omega$. We assume that the wavelength of the electromagnetic
radiation is large compared to the size of the system, so that the
time-dependent field acting on the spins is uniform in space. This
corresponds to situations of practical interest when the microwave
radiation is incident to a thin dielectric layer containing random
magnets. As in microscopic studies of static properties of random magnets \cite{CSS-1986,RA-book,PGC-PRL}, we assume that in the absence of net magnetization the dynamics of the spins is dominated by the local exchange and magnetic anisotropy and neglect the long-range dipole-dipole interaction between the spins. Adding it to the problem would have resulted in a considerable slowdown of the numerical procedure.  

The effective exchange field acting on each spin from the nearest
neighbors in $d$ dimensions is $2dJs$. In our model it competes
with the anisotropy field of strength $2sD_{R}$. The case of a large
random anisotropy, $2sD_{R}\gg2dJs$, that is, $D_{R}\gg dJ$, is
obvious. It corresponds to a system of weakly interacting randomly
oriented spins, each spin aligned with the local anisotropy axis ${\bf n}$.
Due to the two equivalent directions along the easy axis the system
possesses high metastability with the magnetic state depending on
history.

On the contrary, weak anisotropy, $D_{R}\ll dJ$, cannot destroy the
local ferromagnetic order created by the strong exchange interaction.
The direction of the magnetization becomes only slightly disturbed
when one goes from one lattice site to the other. As in the random
walk problem, the deviation of the direction of the magnetization
would grow with the distance. In a $d$-dimensional lattice of spacing
$a$ the average statistical fluctuation of the random anisotropy
field per spin in a volume of size $R$ scales as $D_{{\rm eff}}=2sD_{R}(a/R)^{d/2}$.
Since Heisenberg exchange is equivalent to $J(\nabla{\bf s})^{2}$
in a continuous spin-field model, the ordering effect of the exchange
field scales as $2dJs(a/R)^{2}$. The effective exchange and anisotropy
energies become comparable at $R\sim R_{f}$, where
\begin{equation}
R_{f}\sim a(dJ/D_{R})^{2/(4-d)}\label{Rf}
\end{equation}
 determines the ferromagnetic correlation length. The exact numerical
factor in front of $(J/D_{R})^{2/(4-d)}$ is unknown but the existing approximations
and numerical results suggest that it increases progressively with the dimensionality of
the system \cite{CT-book,PGC-PRL}.

Since magnetic anisotropy has relativistic origin its strength per
spin is usually small compared to the exchange per spin. Anisotropy
axes in the amorphous ferromagnet are determined by the local arrangement
of atoms. When the latter has a short range order the axes are correlated
within structurally ordered grains whose size must replace $a$ in
the RA model. This results in a greater effective RA, making both
limits, $D_{R}\ll dJ$ and $D_{R}\gg dJ$, relevant to amorphous ferromagnets
\cite{CSS-1986,PCG-2015}.

As to the Zeeman interaction of the ac field with the spins, that
is determined by the amplitude of ${\bf h}$ in Eq.\ (\ref{Hamiltonian}), in all situations
of practical interest it would be smaller than all other interactions
by many orders of magnitude. It is worth noticing, however, that for
a sufficiently large system the random energy landscape created by
the RA and the ferromagnetic exchange would have all energy scales,
including that of $h$. This, in principle, may inject nonlinearity
into the problem with a however small amplitude of the ac field.

The undamped dynamics of the system is described by the Larmor equation
\begin{equation}
\hbar\dot{\mathbf{s}}_{i}=\mathbf{s}_{i}\times{\bf H}_{{\rm eff},i},\qquad{\bf H}_{{\rm eff},i}\equiv-\frac{\partial\mathcal{H}}{\partial\mathbf{s}_{i}}\label{Larmor}
\end{equation}
with $\mathbf{h}(t)={\bf h}_{0}\sin(\omega t)$. Here, one also can
include in the usual way \cite{CT-book} a small phenomenological damping $\alpha\ll1$ by adding the term $-\alpha {\bf s}_{i} \times \left(\mathbf{s}_{i}\times{\bf H}_{{\rm eff},i}\right)$ to the first of Eqs.\ (\ref{Larmor}). However, as will be discussed later, a random magnet has a continuous distribution of resonances
(normal modes) that makes the power absorption insensitive to the
small damping. Using the Larmor equation, one can obtain
the relation
\begin{equation}
\dot{\mathcal{H}}(t)=-{\dot{\mathbf h}}(t)\cdot\sum_{i}\mathbf{s}_{i}(t),\label{Absorption}
\end{equation}
where the left-hand side is the rate of change of the total energy (power absorption)
and the right-hand side is the work of the ac field on the magnetic
system per unit of time. For a conservative system this gives two
ways of calculating the absorbed power numerically, which is important for checking
the self-consistency and accuracy of computations. 

When the phenomenological
damping is included, the energy of the spin system saturates in a stationary
state in which the power absorption is balanced by dissipation. On the contrary, the work done by the ac field continues to increase linearly period
after period of the ac field. In this case the work done by the ac field becomes the single measure of the
absorbed power. For a large system, a long computing
time is needed to reach saturation, especially when the damping is small. Fortunately, in most cases the absorbed power can be already obtained with  a good accuracy from a short computation on a conservative system, typically using 5 periods
of the ac field. 

One other complication arises from the necessity to keep the amplitude of the ac field as small
as possible in relation to the exchange and anisotropy to reflect situations of practical interest. In this case, one can expect
normal modes to respond to the ac field independently. However, decreasing the amplitude of the ac field below a certain threshold increases computational errors. Large amplitude of the ac field causes resonant group of the normal
modes to oscillate at higher amplitudes, which triggers nonlinear processes of the energy conversion. 

We use the following procedure. At the first stage, the magnetic state
in zero field is prepared by the energy minimization starting from
random orientation of spins. This reflects the process of manufacturing of amorphous ferromagnet by a rapid freezing from the paramagnetic state in the melt. The numerical method \cite{DCP-PRB2013} combines
sequential rotations of spins ${\bf s}_{i}$ towards the direction
of the local effective field, ${\bf H}_{{\rm eff},i}$, with the probability
$\eta$, and the energy-conserving spin flips (overrelaxation), ${\bf s}_{i}\to2({\bf s}_{i}\cdot{\bf H}_{{\rm eff},i}){\bf H}_{{\rm eff},i}/H_{{\rm eff},i}^{2}-{\bf s}_{i}$,
with the probability $1-\eta$. We used $\eta=0.03$ that ensures
the fastest relaxation. At the end of this stage, a disordered magnetic
state with the ferromagnetic correlation length $R_{f}$ is obtained (see Fig. \ref{Fig-IM}).

At the second stage, the ac field is turned on and Eq. (\ref{Larmor})
is solved with the help of the classical fourth-order Runge-Kutta
method. We also have tried the 5th order Runge-Kutta method by Butcher \cite{Butcher}
that makes six function evaluations per time step. This method can
be faster as it allows a larger time step for the same accuracy. However,
for the RA model it shows instability and has been discarded.
The main computation was parallelized for different ac frequencies
$\omega$ in a cycle over 5 periods of the ac field. Long computation
was required at low frequencies. Wolfram Mathematica with
compilation on a 20-core Dell Precision Workstation was used. In the computations,
we set $s=\hbar=J=1$. In most cases the integration step was  $\Delta t=0.1$
or 0.05. We computed the absorbed power in 1D, 2D, and 3D systems
with a number of spins $N$ up to $400,000$. 

In three dimensions, the ferromagnetic correlation length at small $D_{R}$ is very large and can easily
become longer than the system size. In this case, the magnetization
per spin $m=\left|\mathbf{m}\right|$, where $\mathbf{m}\equiv\left(1/N\right)\sum_{i}\mathbf{s}_{i}$
is the average spin polarization, may be far from zero at the energy minimum. For the sake of uniformity, the ac field was always applied
in the direction perpendicular to $\mathbf{m}$ for which the power
absorption is stronger than in the direction parallel to $\mathbf{m}$.
The absorbed power $P_{abs}$ was obtained by numerically integrating
the left or right side of Eq. (\ref{Absorption}) during $N_{T}$ periods of the ac field and dividing the result by $N_{T}T$ with
$T=1/f=2\pi/\omega$. In most cases we used $N_{T}=5$. It was numerically confirmed
that $P_{abs}\propto h_{0}^{2}$, so, in the plots we show $P_{abs}/h_{0}^{2}$
per spin.

\section{Numerical results}

\label{results} To test our short-time method of computing the absorbed
power, we performed longer computations and plotted the absorbed energy
vs time for the integer number of periods of the ac field, $t=nT$,
$n=0,1,2,\ldots$ For the undamped model, $\alpha=0$, both methods
of computing the absorbed energy discussed in the previous section give the same result, which proves sufficient
computational accuracy. For the damped model, the energy of the system
saturates at long times while the magnetic work, obtained by integrating
the right-hand side of Eq. (\ref{Absorption}) continues to increase
linearly. 

\begin{figure}[h]
\centering{}\includegraphics[width=9cm]{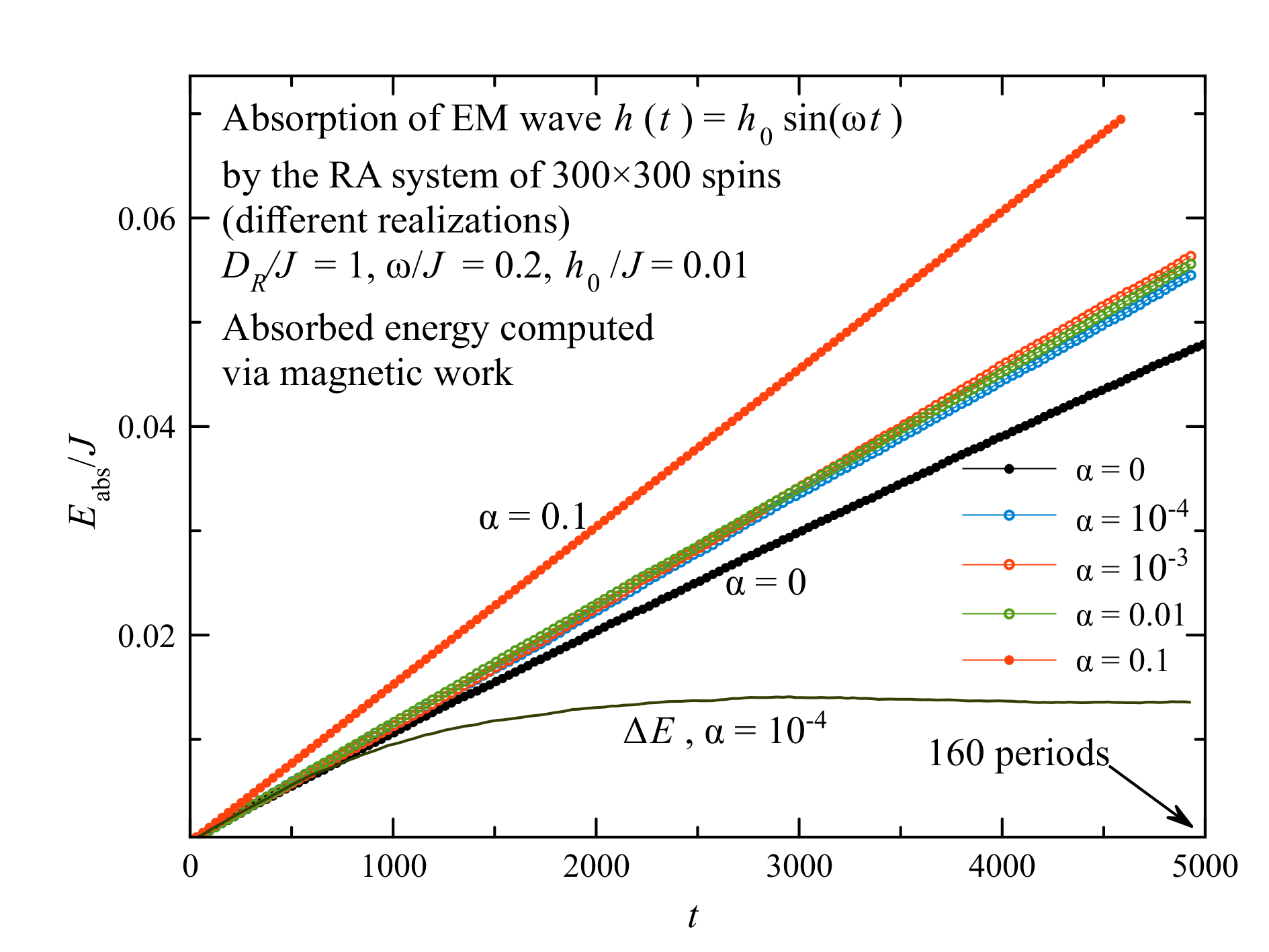} \caption{Absorbed power vs time for a 2D system of $300\times300$ spins for
different values of the damping constant. The energy increase
$\varDelta E$ of the magnetic system caused by the ac power absorption, that saturates at long times due to dissipation, is shown at $\alpha=10^{-4}$. The short-time dependence of the absorbed power is the same
in all cases except for that of the largest damping.}
\label{Fig-2DE-vs-t}
\end{figure}

An example of these tests is shown in Fig. \ref{Fig-2DE-vs-t}
for a 2D system of $300\times300$ spins with $D_{R}/J=1$, $\hbar\omega/J=0.2$,
the ac-field amplitude $h_{0}/J=0.01$, and different values of the
damping constant $\alpha$. For the unrealistically high damping $\alpha=0.1$
the absorbed energy line goes higher than the other dependences. For
$\alpha=10^{-2}$, $10^{-3}$, and $10^{-4}$ the magnetic work is practically the same. This confirms our conjecture about a continuous
distribution of resonances for which the absorption does not depend
of the resonance linewidths, see the next section. In the undamped case, $\alpha=0$, the
absorbed energy goes lower at large times which indicates saturation
of resonances at a given amplitude of the ac-field. 

The plot in Fig. \ref{Fig-2DE-vs-t} at $\alpha=10^{-4}$ shows the increase of the energy of the system, $\Delta E$, for the damping 
that saturates after about 100 periods of the ac field. Such a
long computing time would be impractical  for the computation of the dependence of the
absorbed power on frequency and RA strength. Fortunately, the short-time behavior
of the absorbed energy is the same in all cases, except for the extremely high damping at $\alpha=0.1$.
This allowed us to compute the frequency dependence of the absorbed power by using the undamped model and the short computing time $t=N_{T}T$ with
$N_{T}=5$ or 10. Still, the computation for a large system and low
frequency was rather long.

The choice of the amplitude of the ac-field is important for numerical work. Large
$h_{0}$ reduces numerical noise while leading to the flattening of the
absorption maxima due to a partial saturation. Small $h_{0}$ increases
computational errors. We have chosen the values $h_{0}/J=0.0001$
in 1D (where numerical errors are the smallest), $h_{0}/J=0.0003$ in
2D,  and  $h_{0}/J=0.003$  in 3D.

\begin{figure}[h]
\centering{}\includegraphics[width=9cm]{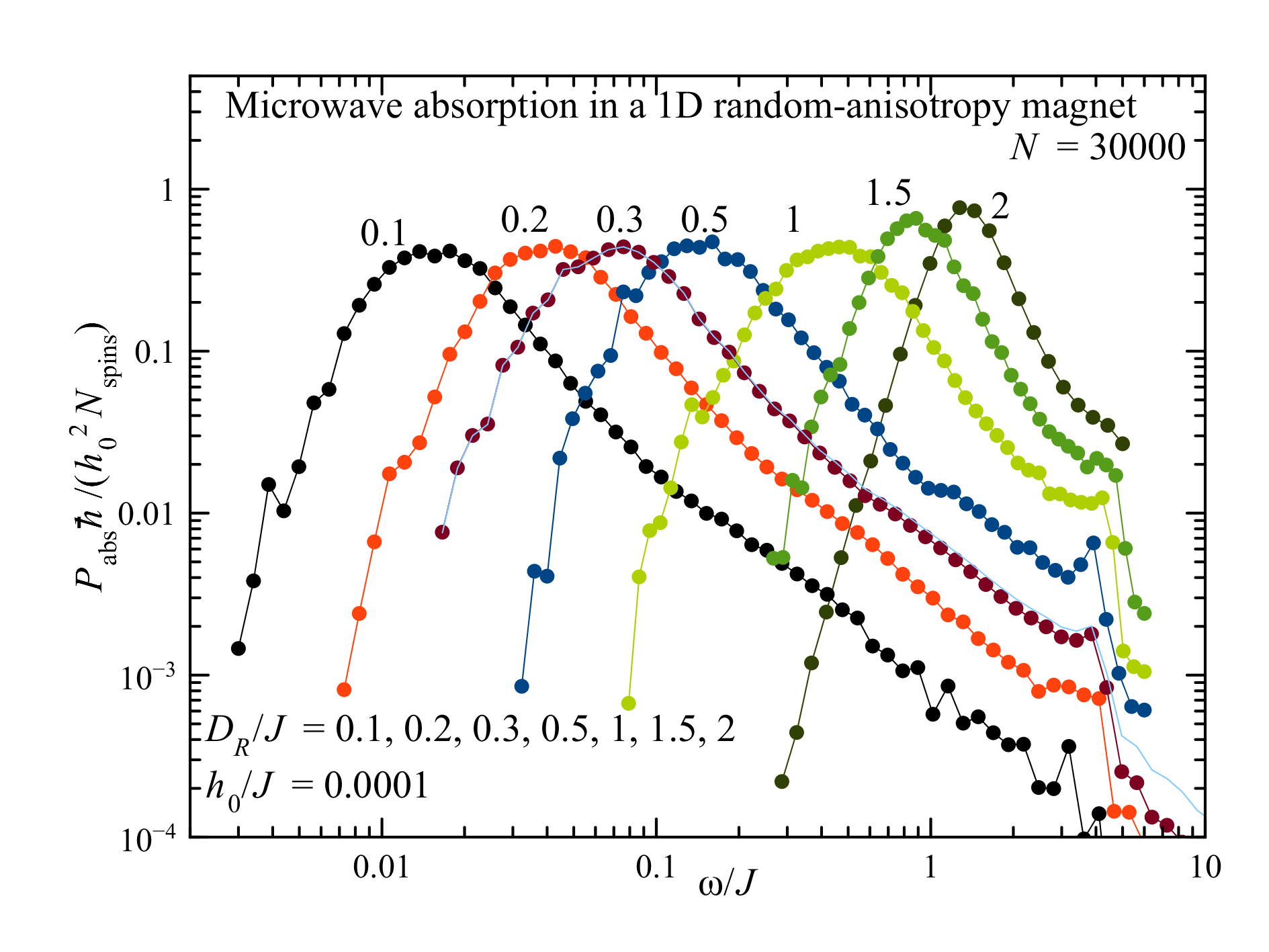} \caption{Absorbed power vs $\omega$ for different values of the random anisotropy
$D_{R}$ in 1D.}
\label{Fig-1DP}
\end{figure}

The frequency dependence of the absorbed power for in a 1D RA ferromagnet is shown
in Fig. \ref{Fig-1DP}. This is the easiest case computationally.
One can use a long chain of spins (here $N=30000$) that is much longer
than the magnetic correlation radius $R_{f}$ in 1D. Thus, after the
energy minimization the system remains well disordered, $m\ll1$.
Fig. \ref{Fig-1DP} shows broad absorption maxima shifting to lower
frequencies with decreasing $D_{R}$. The heights of the maxima are
approximately the same. At large frequencies, there is a cut-off at
the highest spin-wave frequency, $\omega_{\max}=4dJ\Rightarrow4J$
in 1D.

\begin{figure}[h]
\centering{}\includegraphics[width=9cm]{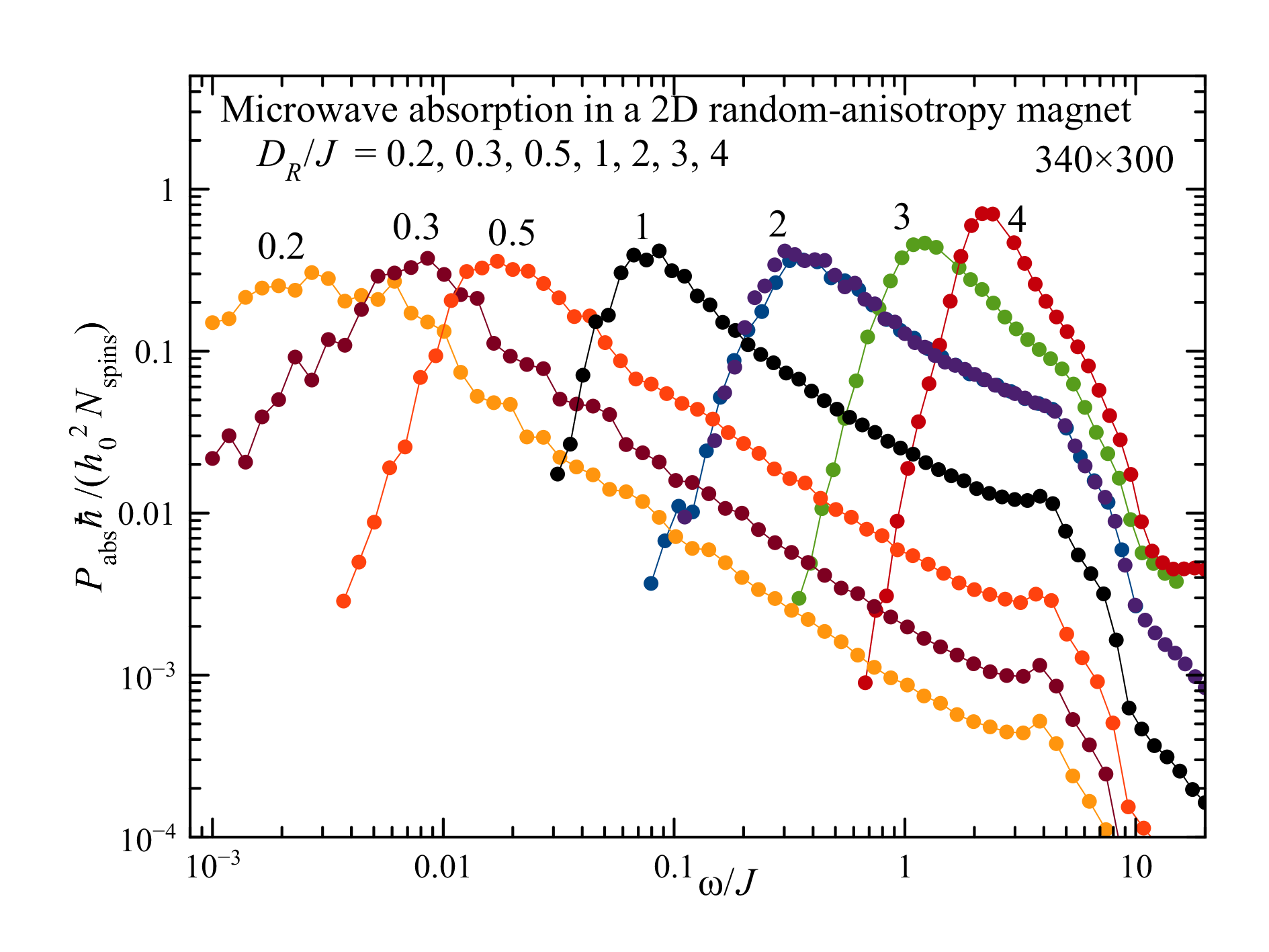} \caption{Absorbed power vs $\omega$ for different values of the random anisotropy
$D_{R}$ in 2D.}
\label{Fig-2DP}
\end{figure}

Frequency dependence of the absorbed power for the 2D model is shown
in Fig. \ref{Fig-2DP}. Qualitatively the results are the same as
in 1D, only we used a larger system of $N=340\times300=102000$ spins.
Here, we were able to go down to the RA only as small as $D_{R}/J=0.2$
as compared to 0.1 in 1D because the absorption maximum is shifting to very
low frequencies on decreasing RA.

\begin{figure}[h]
\centering{}\includegraphics[width=9cm]{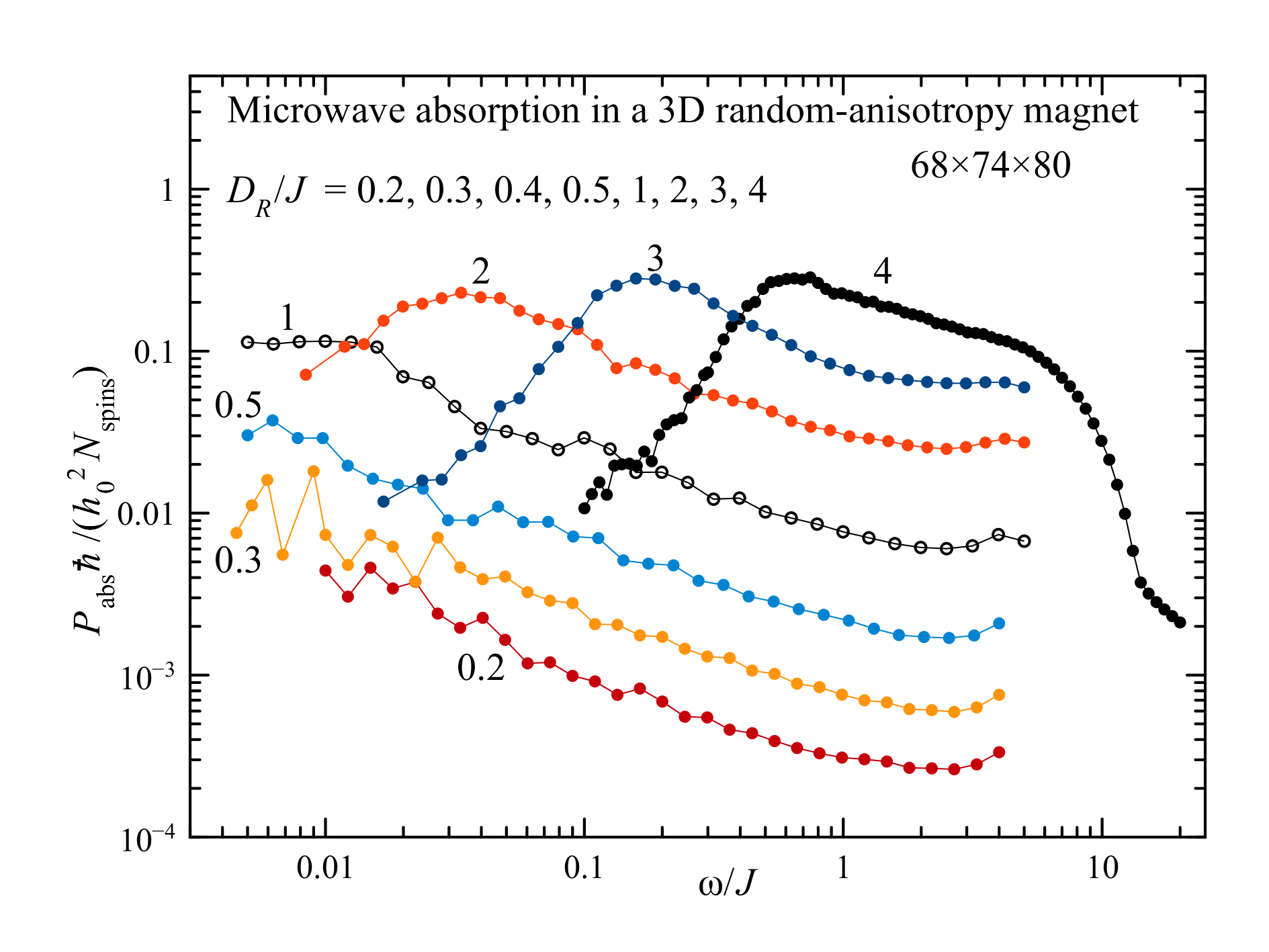} \caption{Absorbed power vs $\omega$ for different values of the random anisotropy
$D_{R}$ in 3D.}
\label{Fig-3DP}
\end{figure}

Frequency dependence of the absorbed power for the 3D model is shown
in Fig. \ref{Fig-3DP}. Again, the absorption curves
are similar to 1D and 2D. However, the 3D model is the hardest
to crack numerically because the ferromagnetic correlation length $R_{f}$, given by
Eq. (\ref{Rf}) with $d=3$, becomes very large at small $D_{R}$.
We had to use 3D systems of a much greater number of spins, $68\times74\times80=402560$
and $75\times80\times84=504000$ , but of smaller lateral dimensions than 1D and 2D systems that we have studied. The lowest RA for which we could observe the absorption maximum in 3D was $D_{R}/J=2$.
For lower $D_{R}$ the absorption maxima shift to very low frequencies
for which computation becomes impractically long and inhibited by the accumulation of numerical errors.

\begin{figure}[h]
\centering{}\includegraphics[width=9cm]{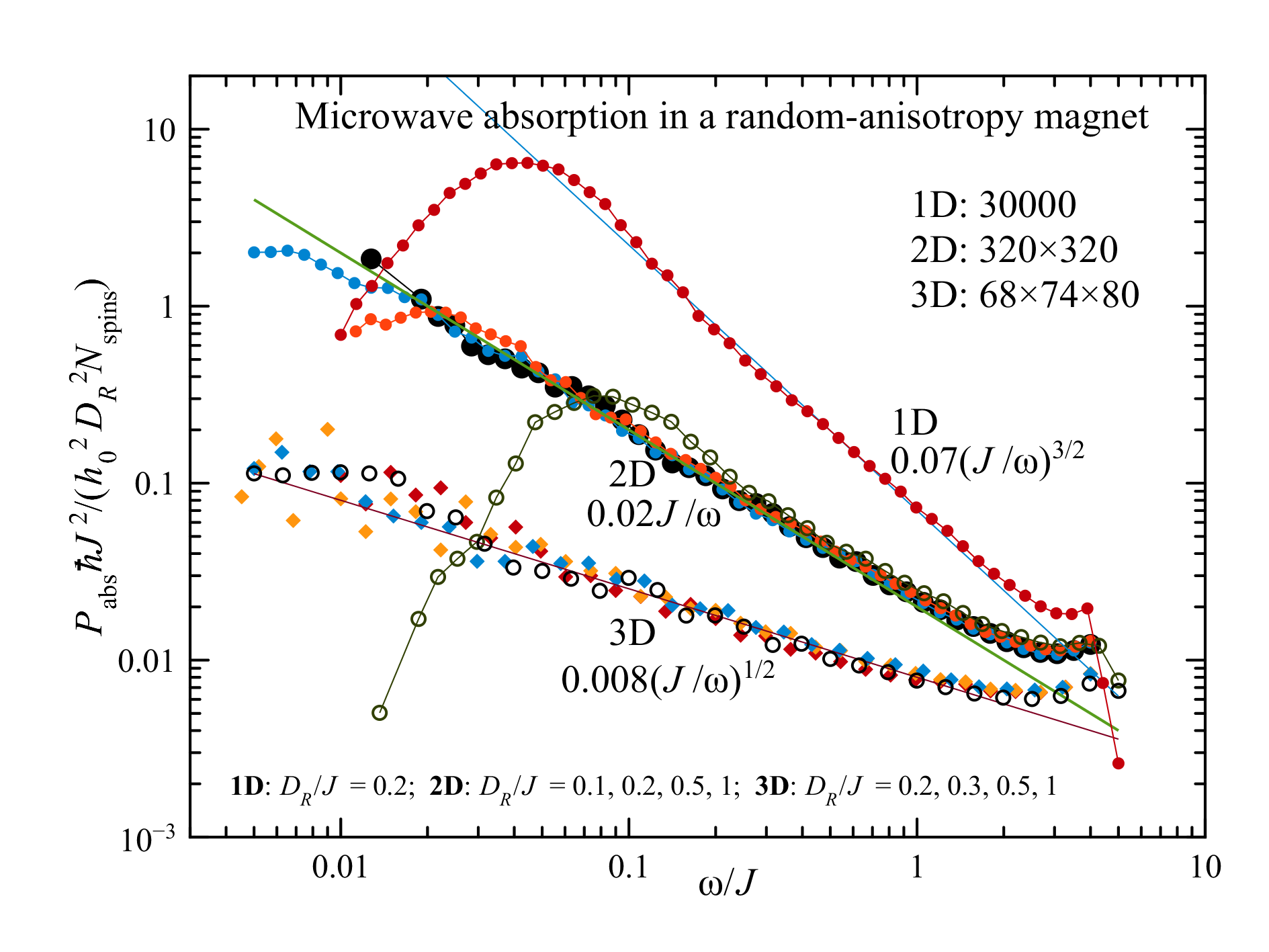} \caption{Scaling representation of the absorbed
power at high frequencies for $d=1,2,3$.}
\label{Fig-All-D-P-scaling}
\end{figure}
Frequency dependence of the power on the right side of the absorption maximum allows scaling shown in Fig. \ref{Fig-All-D-P-scaling}. We have found that $P(\omega)$ in this region follows the power law: 
\begin{equation}
P_{abs}\propto\frac{D_{R}^{2}}{J^{2}}\left(\frac{J}{\hbar\omega}\right)^{\left(4-d\right)/2}. \label{P-omega}
\end{equation}
up to the high-frequency cutoff determined by the strength of the exchange interaction. Away from the maximum the absorption in this high-frequency region is lower in higher dimensions. However the heights of the absorption maxima are comparable in all dimensions, see figures \ref{Fig-1DP}, \ref{Fig-2DP}, and \ref{Fig-3DP}. The maximum absorption has weak dependence on the strength of the RA and the strength of the exchange interaction. By order of magnitude it is given by $P_{\rm max} \sim {h}_{0}^{2}N/\hbar$. 

\section{Interpretation of the results}

\label{interpretation}

\subsection{Independence of the phenomenological damping}

One remarkable feature of the ac power absorption by the RA magnet
is its independence of the damping within a broad range of the damping
constant $\alpha\ll1$, see Fig.\ \ref{Fig-2DE-vs-t}. While we do
not have a rigorous theory that explains this phenomenon, it can be
understood along the lines of the qualitative argument presented below.

The absorption power by a conventional ferromagnet near the FMR frequency,
$\omega\approx\omega_{0}$, has a general form \cite{CT-book}
\begin{equation}
P({\omega,\omega_{0},\alpha})\propto h_{0}^{2}\omega_{0}^{2}\frac{\alpha\omega_{0}^{2}G}{(\omega^{2}-\omega_{0}^{2})^{2}+(\alpha\omega_{0}^{2}G)^{2}}
\end{equation}
with $G$ being a geometrical factor depending on the polarization
of the ac-field and the structure of the magnetic anisotropy. At $\alpha\rightarrow0$
it becomes
\begin{equation}
P({\omega,\omega_{0}})\propto h_{0}^{2}\omega_{0}^{2}\delta(\omega^{2}-\omega_{0}^{2})\propto h^{2}\omega_{0}\delta(\omega-\omega_{0}).
\end{equation}
In an amorphous ferromagnet one should expect many resonances characterized
by some distribution function $f(\omega_{0})$ satisfying
\begin{equation}
\int d\omega_{0}f(\omega_{0})=1.\label{norm-f}
\end{equation}
For the power absorption at a frequency $\omega$ one obtains
\begin{equation}
P(\omega)\propto\int d\omega_{0}f(\omega_{0})h_{0}^{2}\omega_{0}\delta(\omega-\omega_{0})=h_{0}^{2}\omega f(\omega) \label{P-f}
\end{equation}
which is independent of $\alpha$. The function $f(\omega_{0})$ for the RA magnet is unknown. It is related to a more general poorly understood problem of excitation spectrum of systems characterized by a random potential landscape that we are not attempting to solve here. Based upon our numerical results, an argument can be made, however, that sheds light on the physics of the absorption by the RA magnet, see below.

\subsection{Estimation of the maximum-absorption frequency}

\begin{figure}[h]
\centering{}\includegraphics[width=9cm]{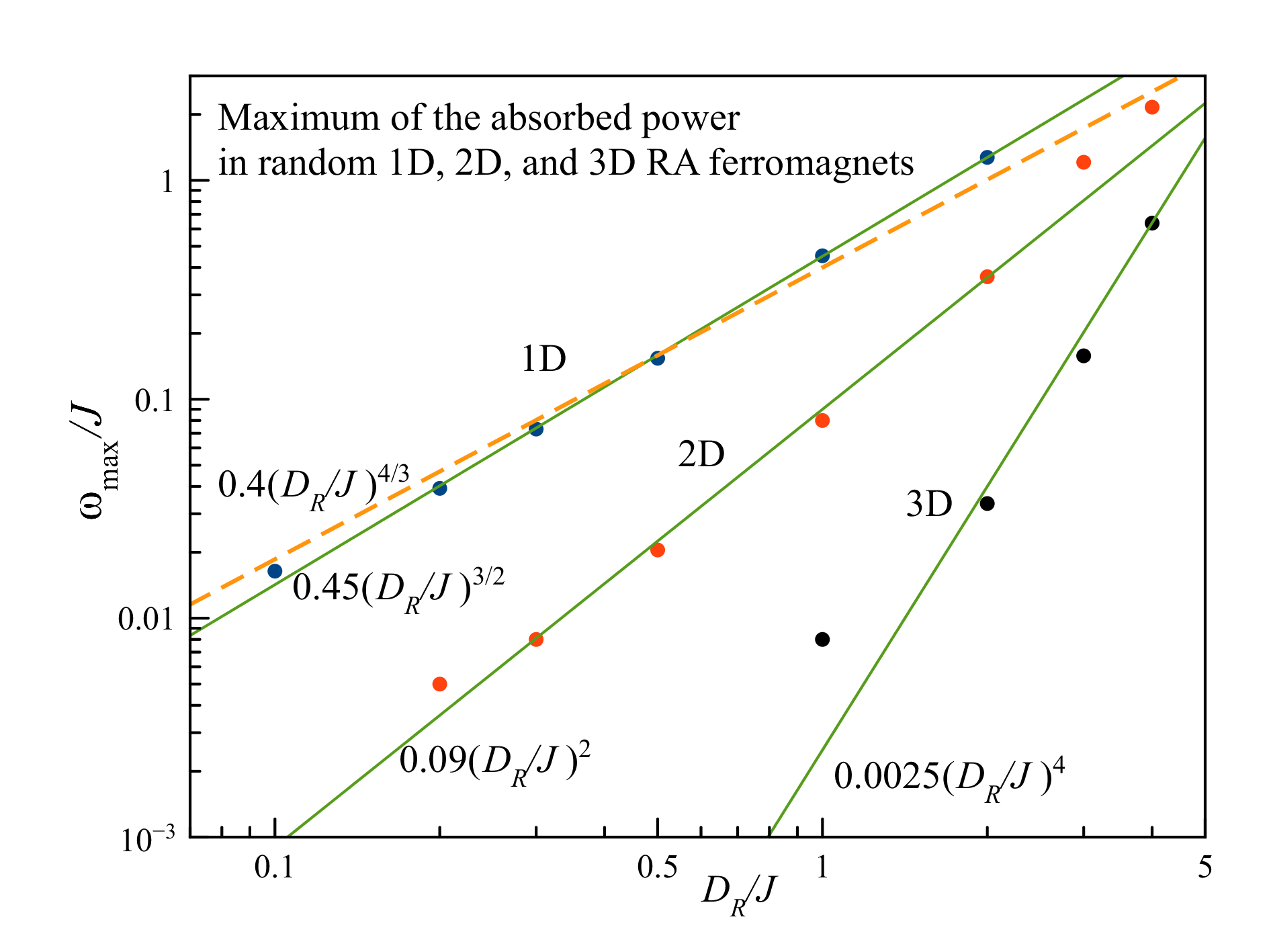} \caption{Peak absorption frequency vs strength of the RA. Points: numerical
experiment. Lines: Power-law fits. In 1D, the expected power law 4/3
is shown by the dashed line while the best fit 3/2 is shown by a solid
line.}
\label{Fig-Max-vs-RA}
\end{figure}

The spin field in the RA ferromagnet resembles to
a some degree a domain structure or magnetization of a sintered magnet comprised of
densely packed single-domain magnetic particles, see Fig.\ (\ref{Fig-IM}). The
essential difference is the absence of boundaries between IM domains.
They are more of a reflection of the disordering on the scale $R_{f}$
than the actual domains. If one nevertheless thinks of the IM domains
as independent ferromagnetically ordered regions of size $R_{f}$,
their FMR frequencies, in the absence of the external field, would
be dominated by the effective magnetic anisotropy, $D_{{\rm eff}}$,
due to statistical fluctuations in the distribution of the RA axes.
In this case the most probable resonance frequency that determines the maximum
of $P(\omega)$ must be given by
\begin{equation}
\omega_{{\rm max}}\sim D_{{\rm eff}}\sim D_{R}(a/R_{f})^{d/2}. \label{omega-R}
\end{equation}
Substituting here $R_{f}/a\sim k(d)(J/D_{R})^{2/(4-d)}$, with the
factor $k(d)$ increasing \cite{CSS-1986} progressively with $d$,
we obtain
\begin{equation}
\frac{\omega_{{\rm max}}}{J}=\frac{1}{k^{d/2}}\left(\frac{D_{R}}{J}\right)^{4/(4-d)}.\label{omega-max}
\end{equation}
It suggests that ${\omega_{{\rm max}}}/{J}$ must scale as $(J/D_{R})^{4/3}$
in one dimension, as $(J/D_{R})^{2}$ in two dimensions, and as $(J/D_{R})^{4}$
in three dimensions.

The dependence of ${\omega_{{\rm max}}}/{J}$ on $J/D_{R}$ for
$d=1,2,3$ derived from figures \ \ref{Fig-1DP}, \ref{Fig-2DP},
and \ref{Fig-3DP} is shown in Fig. \ref{Fig-Max-vs-RA}. In 2D and
3D there is a full agreement with the above argument. In
1D the best fit seems to be the $3/2$ power of $J/D_{R}$ instead of the expected $4/3$ power. Given
the qualitative nature of the argument presented above and good fit for $d=2,3$ the agreement
is nevertheless quite good. The small factor in front of the power of $J/D_{R}$, that becomes
progressively smaller as one goes from $d=1$ to $d=3$, correlates
with the established fact \cite{CSS-1986,DC-1991} that $k(d)$
increases with $d$.

\begin{figure}[h]
\centering{}\includegraphics[width=9cm]{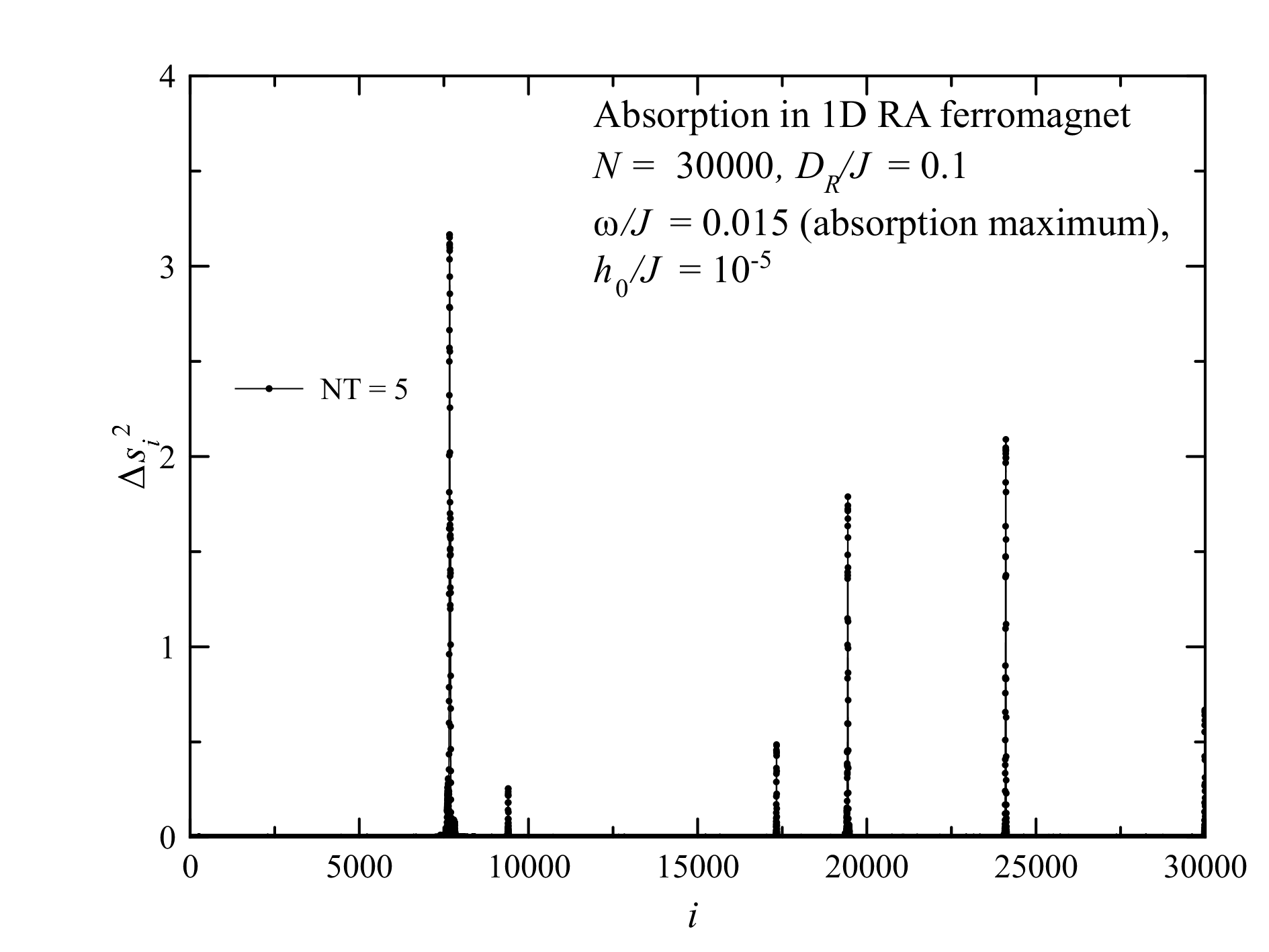} \caption{Spatial dependence of spin deviations in a 1D RA
ferromagnet.}
\label{fig8}
\end{figure}

\subsection{Vizualization of the local dynamical modes}

Further evidence of the validity of our picture that the power absorption occurs inside
resonant IM domains comes from the analysis of the spatial dependence of local spin deviations from the initial state $s_i^{(0)}$, defined as $\Delta s_i^2 = (s_i -s_i^{(0)})^2$. They are related to local spin oscillations and are illustrated in Fig.\ \ref{fig8} for a 1D RA ferromagnet. Noticeable deviations
occur at discrete locations. Their amplitude is apparently determined by how well the frequency of the ac field matches the resonant frequency
of the IM domain at that location. The oscillating domains appear to
be well separated in space.

\begin{figure}[h]
\centering{}\includegraphics[width=9cm]{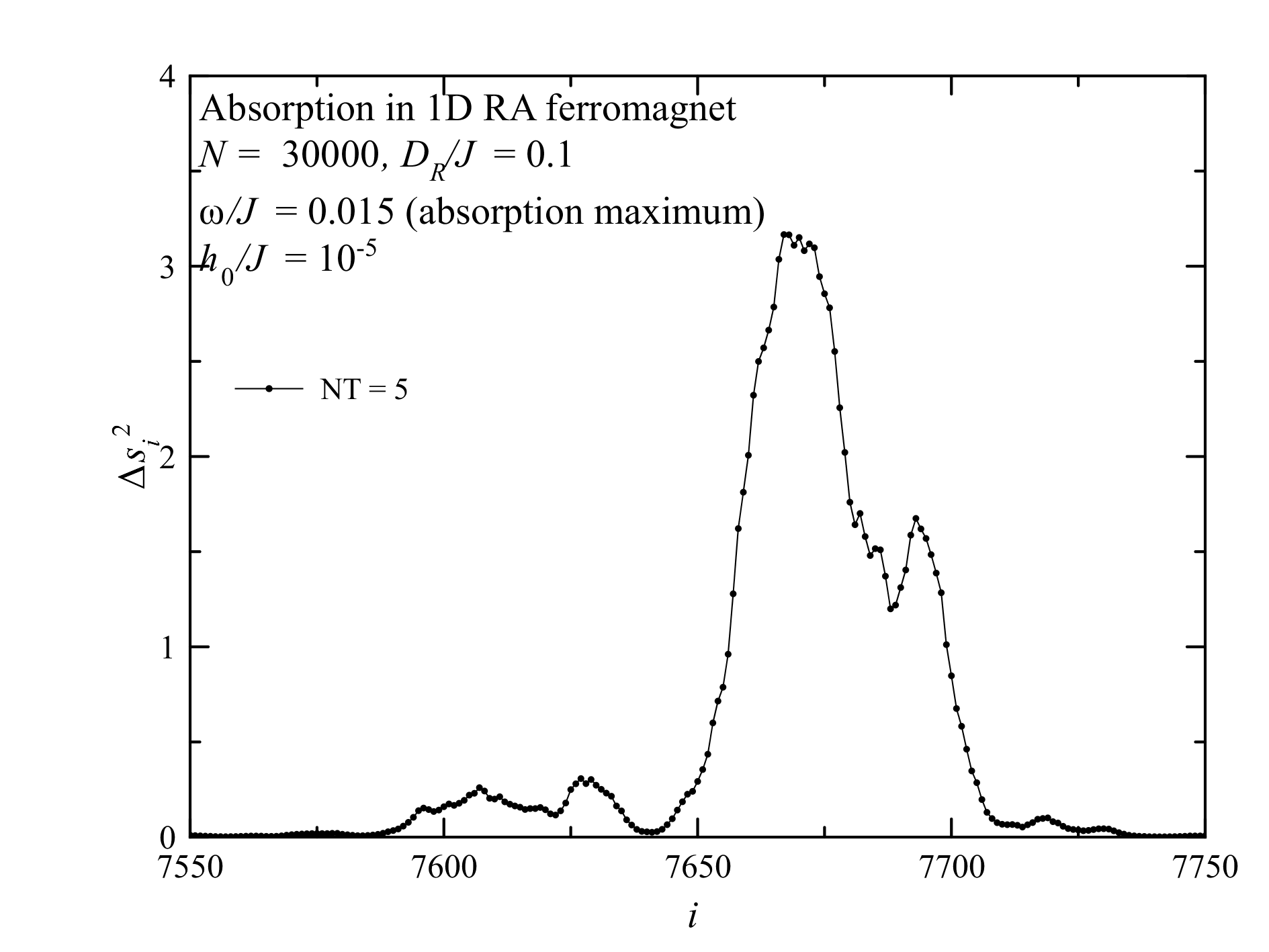} \caption{Spatial dependence of spin deviations in one area
of resonant absorption.}
\label{fig9}
\end{figure}

One of the oscillating regions of Fig. \ref{fig8} is zoomed at in
Fig.\ \ref{fig9}. It shows that the oscillations quickly go to zero
away from the region. The width of the region correlates with the
expected value of $R_{f}$.

\begin{figure}[h]
\centering{}\includegraphics[width=9cm]{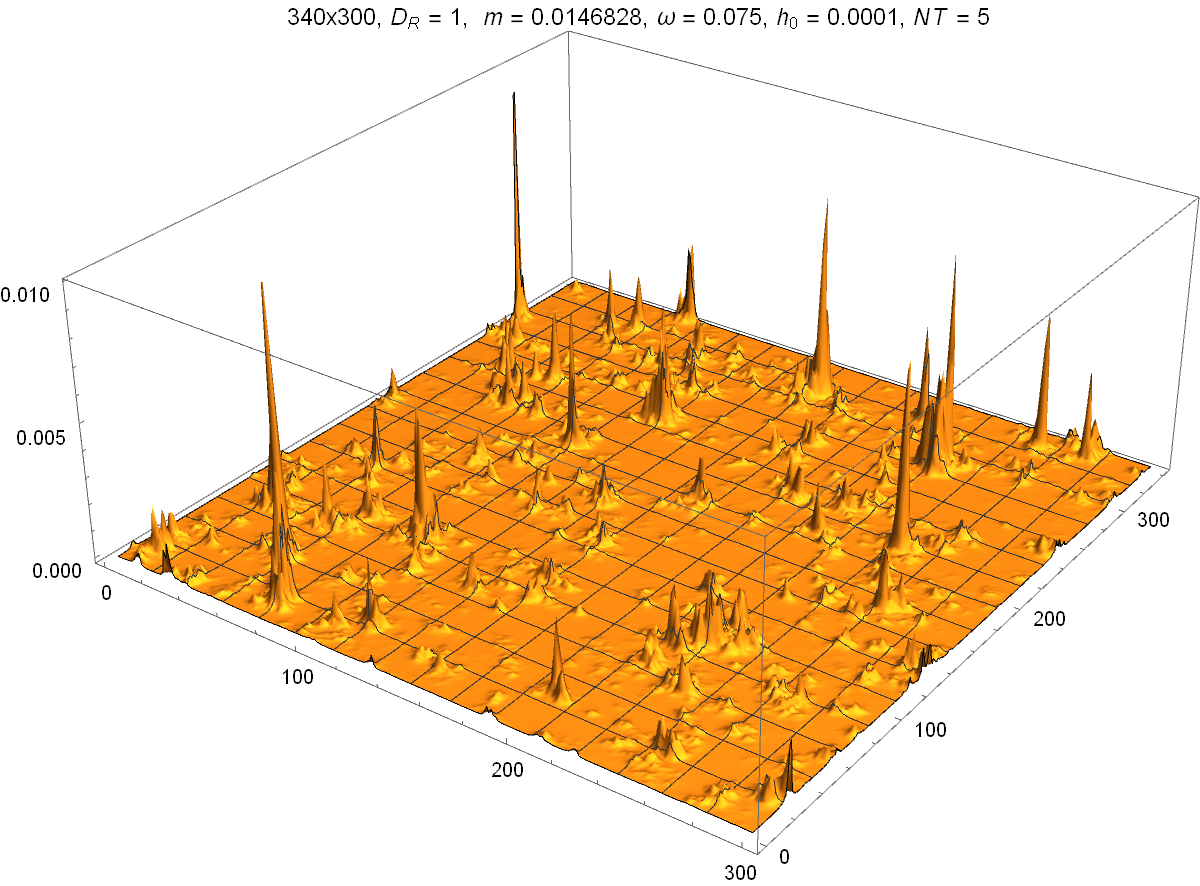} \caption{Spatial dependence of spin deviations in a 2D RA
ferromagnet.}
\label{fig10}
\end{figure}

Fig.\ \ref{fig10} shows oscillating regions in a 2D RA ferromagnet.
Here again the spin regions that absorb the ac power are well separated
in space. This is in line with our picture that they correspond to
the resonant IM domains in which the effective magnetic anisotropy
due to statistical fluctuations of easy axis directions matches the
frequency of the ac field. 

Note that the peaks in figures \ref{fig8}, \ref{fig9}, and \ref{fig10} are not stationary, they go up and down in time, but their locations in space are fixed. The snapshots shown in these figures have been taken at particular moments in time. 

At present we do not have the theory of the full dependence of the power
on frequency. Apparently, it is related to the size distribution of
ferromagnetically correlated regions (Imry-Ma domains), which remains
a challenging unsolved problem of statistical mechanics. Our numerical findings, however, may shed some partial light onto this problem.

Indeed, we have found that at large frequencies the power absorption follows Eq.\ (\ref{P-omega}): $P \propto \omega^{(d/2) - 2}$. If it is related to the precession of IM domains of size $R$, then according to Eq.\ (\ref{omega-R}) the frequency of this precession scales as $\omega \propto R^{-d/2}$. Since this is a high-frequency regime, it must correspond to small $R$. According to Eq.\  (\ref{P-f}) $P \propto \omega f(\omega)$. Combined with Eq.\ (\ref{P-omega}) it gives  $f(\omega)  \propto \omega^{(d/2) - 3}$ at large $\omega$. If distribution of IM domains is given by $F(R)$ satisfying $\int dR F(R) =\int d\omega f(\omega)=1$, then, using the above formulas we obtain 
\begin{equation}
F(R) = f(\omega)\frac{d\omega}{dR} \propto R^{-\left(\frac{d}{2} - 1\right)^2}. \label{F-R}
\end{equation}
This suggests $R^{-1/4}$ distribution of small-size IM domains in 1D and 3D, and independence of $R$ (up to a log factor) in 2D. 

The qualitative argument leading to Eq.\ (\ref{F-R}) is based upon the picture of independently oscillating IM domains. In reality there are no boundaries between ferromagnetically correlated regions. Our derivation suggests a large fraction of compact correlated regions of size that is small compared to the ferromagnetic correlation length $R_f$. It is supported by Fig.\ \ref{Fig-IM} but is different from the prediction of the exponentially small number of such regions in the random-field xy model made within variational approach \cite{Garel}. 

\section{Discussion}

\label{discussion}

We have studied the power absorption by the random-anisotropy ferromagnet
in a microwave field in one, two, and three dimensions. The one-dimensional
problem describes a thin wire of diameter smaller than the 1D ferromagnetic
correlation length $R_{f}$ and of length greater than $R_{f}$. The
two-dimensional problem corresponds to a film of thickness that is
small compared to the 2D ferromagnetic correlation length and of lateral
dimension large compared to $R_{f}$. The three-dimensional problem
corresponds to a particle of amorphous ferromagnet of size large compared
to the 3D ferromagnetic correlation length.

Our main finding agrees with the statements made by experimentalists \cite{Suran-localization}. It elucidates the physics of the microwave absorption
by an RA ferromagnet. The absorption is localized inside
well separated regions. Scaling of the peak absorption frequency with
the strength of the RA points towards the mechanism of the absorption in
which oscillations of spins are dominated by isolated ferromagnetically
correlated regions (Imry-Ma domains) that are in resonance with the
microwave field. 

Broad distribution of sizes of ferromagnetically correlated regions results in the broad distribution of resonance frequencies. It makes the absorption broadband, with the frequency at half maximum covering two orders of magnitude. Another consequence of such a broad distribution is independence of the absorption on the damping of spin oscillations within a few orders of magnitude of the damping constant. 

A remarkable observation is that the maximum of the absorbed power in a random magnet has a weak dependence on basically all parameters of the system, such as dimensionality, damping, the strength of the RA, and the strength of the exchange interaction.  By the order of magnitude it is determined solely by the total number of spins absorbing the microwave energy and the amplitude of the microwave field. This again is a consequence of the broad distribution of sizes of ferromagnetically correlated regions, causing broad distribution of the effective magnetic anisotropy and effective exchange interaction.

A practical question is whether the RA (amorphous) magnets have a good prospect
as microwave absorbers. The power absorption by the random magnet occurs in the broad frequency range. Frequencies that provide the maximum of the absorption depend on the strength of the RA. The latter can be varied by at least two orders of magnitude by choosing soft or hard magnetic materials in the process of manufacturing an amorphous magnet. It must allow the peak absorption in the range from a few GHz to hundreds of GHz. 

As we have already mentioned, the power absorption at frequencies near the absorption maximum depends weakly on the parameters of the system. By order of magnitude it equals
\begin{equation}
P_{\rm max} \sim {h}_{0}^{2}N/\hbar= 4\mu_{B}^{2}S^{2}B_{0}^{2}n_{0}Ad/\hbar. \label{P-estimate}
\end{equation}
Here we have written ${h}_{0}=2\mu_{B}SB_{0}$ in terms of the length of the dimensionless spin $S$ and the dimensional amplitude, $B_0$, of the microwave field, with $\mu_B$ being the Bohr magneton. The number of spins in the absorbing layer, $N = n_0 Ad$, has been expressed in terms of the concentration of spins $n_0$, the area $A$ and the thickness $d$ of the layer. 

The incoming microwave delivers to the layer the power $P_{m}=({cB_{0}^{2}}/{2\mu_{0}})A$,
where $\mu_{0}$ is the permeability of free space \cite{Jackson}. This gives for the ratio of the absorbed and incoming powers 
\begin{equation}
\frac{P_{\rm max}}{P_{m}} \sim \frac{8S^{2}\mu_{0}\mu_{B}^{2}n_{0}d}{\hbar c}.
\end{equation}
For a rough estimate, at $S = 2$ and $n_0 \sim 2 \times 10^{27}m^{-3}$ this ratio becomes of order unity at $d \sim 5mm$. It suggests that a layer of such thickness composed of magnetically dense amorphous wires, foils, and particles, thinly coated to prevent reflectivity due to the electric conductance, may be a strong absorber of the microwave radiation. Parameter-dependent numerical factor of order unity that we omitted in Eq.\ \ref{P-estimate} may slightly favor low-dimensinal systems in that respect.
\\

\section{Acknowledgements}

This work has been supported by the grant No. 20RT0090 funded by the
Air Force Office of Scientific Research.

\end{document}